\documentclass[11pt,a4paper]{article}
\pdfoutput=1
\usepackage{jheppub}

\usepackage{amsmath}
\usepackage{verbatim}
\usepackage{amssymb}
\usepackage{inputenc, array}
\usepackage{textcomp}
\usepackage{bm}
\usepackage{physics}
\usepackage{simpler-wick}
\usepackage{appendix}
\usepackage{xcolor}

\newcommand{\e}{\epsilon}
\renewcommand{\d}{\partial}
\newcommand{\be}[1]{\begin{equation}\label{#1} }
\newcommand{\ee}{\end{equation}}
\newcommand{\bea}[1]{\begin{eqnarray}\label{#1} }
\newcommand{\eea}{\end{eqnarray}}
\newcommand{\refb}[1]{(\ref{#1})}

\renewcommand{\O}{{\mathcal{O}}}

\pdfoutput=1
\usepackage{array}
\usepackage{multirow}
\usepackage{amsmath}
\usepackage{makecell}

\definecolor{green}{rgb}{0.1,0.8,0.2}

\usepackage{cancel}
\usepackage{graphicx} % for \includegraphics command and options

\usepackage{hyperref}
\usepackage{cleveref}
\usepackage{float}
\usepackage{subcaption}
\usepackage{physics}
\usepackage{slashed}
\usepackage{amsfonts}
\usepackage{caption}
\usepackage{mathrsfs}

\allowdisplaybreaks
\title{Anatomy of Null Contractions}

\author{Arjun Bagchi,} \author{Nachiketh M,} \author{and Pushkar Soni.}  \author{\\}

\affiliation{Indian Institute of Technology Kanpur, Kanpur 208016, India.\\ } 

\emailAdd{(abagchi, nachikethm22, pushkars21)@iitk.ac.in}

\preprint{}
\abstract{We introduce null contractions of the Poincare and relativistic conformal algebras. The longitudinal null contraction involves writing the algebra in lightcone coordinates and contracting one of the null directions. For the Poincare algebra, this yields {\em two non-overlapping co-dimension one} Carroll algebras. The transverse contraction is a limit on the spatial dimensions and yields two non-overlapping co-dimension one Galilean algebras. We find, similar to Susskind's original observation of the non-relativistic case, that the Poincare algebra, written in the lightcone coordinates, naturally contains Carrollian sub-algebras in one lower dimension. The effect of the longitudinal contraction, which essentially focusses on the null direction, is to disentangle the two Carroll algebras that now correspond to the symmetries of the two null boundaries. The transverse contraction similarly separates the overlapping Galilean sub-algebras of the original Poincare algebra. We discuss aspects of the conformal case, where we get lower dimensional Carroll Conformal algebras and Schr{\"o}dinger algebras.}

\begin{document}
\maketitle

\section{Introduction}
Non-Lorentzian physics has been a topic of a lot of recent interest with novel geometric structures emerging from the usual Riemmannian geometry when the speed of light $c$ is dialled to infinity or zero. We point the reader to the reviews \cite{Bergshoeff:2022eog, Hartong:2022lsy, Oling:2022fft} for some idea of the recent developments in the field. 

\subsection*{The resurgence of Carroll}
The Galilean regime ($c\to \infty$) has always been of interest for various physical applications in the real world where the characteristic velocities are much smaller compared to the speed of light. Of late, the Carroll regime ($c\to0$) \cite{LBLL, SenGupta:1966qer} has become particularly important with very wide ranging applications. Carrollian physics has emerged in condensed matter systems like fractons \cite{Bidussi:2021nmp} and set-ups with flat bands \cite{Bagchi:2022eui}, in hydrodynamics e.g. in ultra-relativistic fluids \cite{Bagchi:2023ysc, Bagchi:2023rwd}. In the context of gravity, Carrollian symmetries have been shown to be important in black holes, e.g. Carroll structures appear on horizons of black holes \cite{Donnay:2019jiz}, and in cosmology \cite{deBoer:2021jej}. In quantum gravity, in perhaps the most important of the applications, (conformal) Carrollian theories have been proposed as duals to asymptotically flat spacetimes \cite{Bagchi:2010zz, Bagchi:2012xr, Bagchi:2016bcd, Bagchi:2022emh,Donnay:2022aba}. Finally, Carrollian symmetry has also been explored in the context of string theory. This includes important questions about string theory in the very high energy/tensionless limit \cite{Bagchi:2013bga, Bagchi:2015nca, Bagchi:2020fpr}, strings probing black holes \cite{Bagchi:2023cfp} and models to black hole microstates \cite{Bagchi:2022iqb}. 

\medskip

Carroll symmetries naturally arise on null surfaces. One of the debates in the community has been associated with the confusion of calling these symmetries ``ultra-local" versus ``ultra-relativistic". Sending the speed of light to zero means the closing up of the light cone and hence the ultra-local nature of the theory is not in doubt. But it has been argued (by one of us in particular) that this should be also thought of as an ultra-relativistic limit, since we could think of the symmetries from a co-dimension one point of view and the theory that lives on a constant time slice by virtue of this limit is boosted to a constant null slice, thus making the symmetries contract from the Poincare to the Carroll algebra. 

\subsection*{Null contractions}

In an effort to clarify the above point and show that the Carroll limit is indeed an ultra-relativistic limit, in this paper, we perform a different contraction, where the parent algebra is written in lightcone coordinates and then one focusses on a light-like direction (we will call this the longitudinal null contraction). We will show that a lower dimensional Carroll algebra arises naturally. Interestingly, and in hindsight very expectedly, we actually get two overlapping co-dimension one Carrollian algebras. We also get a certain boost generator which survives the limit and acts similar to a dilatation operator on the system labelling the generators under boost weights. The reason why this contraction is different from the standard Carroll contraction is that lightcone coordinates cannot be reached by a finite Lorentz transformation from any inertial coordinate system. 

\medskip

Instead of a contraction that focusses on the lightlike coordinate, we could do a transverse null contraction on the spatial directions and this now yields two overlapping co-dimension one Galilean algebras, with again a boost generator. These Galilean algebras can be extended to Bargmann (or massive Galilean) algebras as well. 

\medskip

Our construction is reminiscent of the seminal work by Susskind \cite{Susskind:1967rg}, where he discovered a lower dimensional Galilean sub-algebra inside the Poincare algebra when going to the lightcone coordinates (see \cite{Kogut:1971zd} for a more detailed discussion). We revisit this work and infer that the contraction that we do is precisely the algebraic way of zooming into this subalgebra. We notice that there are two such subalgebras completely in keeping with our analysis. The contraction that we do gets rid of the non-zero cross commutations between these subalgebras and lands us up on Galilean world. 

\medskip

We then observe that the Poincare algebra also admits two Carroll sub-algebras, like the Galilean case found by Susskind and the contraction again separates these two sub-algebras which live on the null surface. 

\subsection*{A conundrum and a solution}
In doing this exercise, we solve a rather intriguing puzzle. It has been long postulated that when an observer is highly boosted, she observes Galilean physics. This is seemingly at odds with the statement that the symmetries associated with null surfaces, and hence with light-like boosted observers, is Carrollian. What our algebraic analysis clarifies is that there are both Carrollian and Galilean symmetries associated with highly boosted observers. When these observers are interested in physics applicable to themselves, this would correspond to focussing on the null directions and hence the physics would be Carrollian. If they are interested in the physics of the transverse directions, i.e. the stuff they are flying past, this would correspond to the transverse contraction and they would observe Galilean physics, in keeping with earlier work. 

\subsection*{Conformal extensions}
We look at the conformal extension of our construction. In the transverse contraction, we find results along expected lines. There are two co-dimension one $z=1$ Carrollian Conformal algebras in the relativistic conformal algebra and in the limit these become disjoint. Here $z$ is the dynamical exponent, which gives the relative scaling between space and time. Relativistic theories of course naturally have $z=1$ because space and time are on the same footing. Interestingly, the non-relativistic case is different. The relativistic conformal algebra does not have the $z=1$ Galilean conformal algebras as sub-algebras, but their $z=2$ variants, the Schr{\"o}dinger algebra. We also observe that the transverse contraction does not get rid of the overlap between the two Schr{\"o}dinger algebras. 

\subsection*{Outline of the paper}
The rest of the paper is organised as follows. In Sec.~2, we recapitulate the basics of the two In{\"o}n{\"u}-Wigner contractions of the Poincare algebra, the Carroll ($c\to0$) and Galilei ($c\to\infty$) and highlight the difference between the two contracted algebras. Sec.~3 is the most important section of the paper where we lay down the details of the null contractions and obtain the lower dimensional Carrollian and Galilean algebras. We rewrite the Poincare algebra in lightcone coordinates, identify the Carroll and Galilean subalgebras and show how the longitudinal and transverse contractions disentangle the overlapping algebras. We end this section with some discussions. Sec.~4 contains the details of the conformal contractions. Here we first remind the reader of the Carrollian conformal algebra and the Galilean versions, the Galilean conformal algebra and the Schr{\"o}dinger algebra. We then perform analogous longitudinal and transverse contractions on the relativistic conformal algebra. We highlight the differences between the Carrollian and Galilean cases. In Sec.~5, we end with discussions and future directions. 

\medskip

\paragraph{Note added:} When this paper was being readied for submission, \cite{Majumdar:2024rxg} appeared on the arXiv, where it was also observed that the lightcone Poincare algebra has lower dimensional Carroll subalgebras. 

\newpage

\section{Contractions: Carroll and Galilei}
In this section, we revisit the now well-known contractions of the Poincare algebra in $d$ spacetime dimensions. The Galilean algebra is obtained by sending the speed of light $c$ to infinity ($c\to \infty$) while the diametrically opposite limit $c\to0$, leads to the Carroll algebra. Below we explicitly obtain both algebras. 

\subsection{Galilean contraction}
The Poincare algebra is given by the Lorentz generators $J_{\mu\nu}$ and translations $P_\mu$ which in Cartesian coordinates are given by 
\be{poincare}
J_{\mu\nu}= x_\mu\partial_\nu - x_\nu\partial_\mu, \quad P_\mu = \partial_\mu
\ee
These close to form the well-know Poincare algebra:
\begin{subequations}
\bea{}
&& [J_{\mu \nu}, J_{\rho\sigma}] = \eta_{\rho \mu} J_{\sigma\nu } - \eta_{\rho \nu} J_{\sigma\mu } + \eta_{\sigma \nu} J_{\rho\mu} - \eta_{\sigma \mu} J_{\rho\nu },  \\
&& [J_{\mu \nu}, P_{\rho}] = \eta_{\rho \nu} P_{\mu}  -  \eta_{\mu \rho} P_\nu, \quad [P_\mu, P_\nu]=0.
\eea
\end{subequations}
The Galilean limit is where we send the speed of light $c$ to infinity. The usual way of doing this is to write $x_0 = ct$ and keep track of factors of $c$ in the generators. We adopt a slightly different but equivalent way of implementing this by doing the following:
\be{}
x_i \to \e x_i, \quad t \to t, \quad \e \to 0.
\ee
Here $\e$ can be thought of as the ratio of a characteristic velocity $v$ and $c$, i.e. $\e = \frac{v}{c}$, hence $\e\to 0$ is equivalent to sending $c\to \infty$ while holding $v$ fixed. In this limit, it is easily shown that the Poincare algebra contracts to the Galilean one. Specifically, consider the boost generator:
\be{}
J_{0i} = t\d_i +x_i \d_t \rightarrow \frac{1}{\e} t \d_i + \e x_i\d_t 
\ee 
In order to keep the generator finite in the limit, we define Galilean boosts: 
\be{}
G_i = \lim_{\e\to0} \e J_{0i} = t \d_i.
\ee
After appropriately regularising the generators in the limit, we get the Galilean generators as follows: 
\be{}
H= \d_t, \,P_i = \d_i, \, G_i = t \d_i, \, J_{ij} = x_i\d_j -x_j\d_i. 
\ee
The commutation relations are
\begin{subequations}\label{gal}
\bea{}
&& [H, G_i] = P_i, \quad [G_i, P_j] = 0, \quad [H, P_i]=0, \quad [P_i, P_j]= 0, \quad [G_i, G_j]=0, \\
&& [P_{k}, J_{ij}] = \delta_{ik} P_j  - \delta_{jk} P_{i}, \quad  [G_{k}, J_{ij}] = \delta_{ik} G_j  - \delta_{jk} G_{i}, \quad [J_{ij}, H] = 0, \\
&& [J_{ij}, J_{kl}] = so(d)
\eea
\end{subequations}
The crucial point of difference with the Poincare algebra is that the boosts $G_i$ now commute. The Galilean algebra can be given a central extension:
\be{barg}
[G_i, P_j] = M\delta_{ij}
\ee
The generator $M$ is central, i.e. it commutes with all generators. This centrally extended Galilean algebra is called the Bargmann algebra and the $M$ plays the role of mass in the non-relativistic system.

\subsection{Carroll contraction}
We will now consider the Carroll limit, where the speed of light is sent to zero. In our dimensionless way of taking the limit, this amounts to 
\be{carr-limit}
x_i \to x_i, \quad t \to \e t, \quad \e \to 0.
\ee
or a contraction of the timelike direction. So the time-direction would now become null. Again, in terms of the generators, the important one is the boost. The Lorentz boost now contracts in the following way
\be{}
J_{0i}  \rightarrow \frac{1}{\e} x_i\d_t  + \e t \d_i.  
\ee 
To keep the generator finite in the Carroll limit, we now define Carrollian boosts as 
\be{}
C_i = \lim_{\e\to0} \e J_{0i} = x_i\d_t.
\ee
The Carrollian generators are given by 
\be{carr-gen}
H= \d_t, \,P_i = \d_i, \, C_i = x_i \d_t, \, J_{ij} = x_i\d_j -x_j\d_i. 
\ee
The commutation relations are
\begin{subequations}\label{carr}
\bea{}
&& [H, C_i] = 0, \quad [C_i, P_j] = - H \delta_{ij}, \quad [H, P_i]=0, \quad [P_i, P_j]= 0, \, \, [C_i, C_j]=0, \\
&& [J_{ij}, P_{k}] =  \delta_{k[j} P_{i]} \quad  [J_{ij}, C_{k}] = \delta_{k[j} C_{i]}, \quad [J_{ij}, H] = 0, [J_{ij}, J_{kl}] = so(d).
\eea
\end{subequations}
Note that in the Carroll algebra, the Hamiltonian has become a central element, but boosts still commute as in the Galilean case. It is important to highlight the points of difference between the Galilean and Carrollian algebras. This is of course to do with the different boost generators. We explicitly write this below: 
\bea{}
&&\text{Galilei:} \quad [H, G_i] = P_i, \quad [G_i, P_j] = 0. \label{gal-def}\\
&&\text{Carroll:} \quad [H, C_i] = 0, \quad [C_i, P_j] = - H \delta_{ij}. \label{car-def}
\eea
The Bargmann algebra additionally has $[G_i, P_j] =M\delta_{ij}$, but the Bargmann algebra and the Carroll algebra are not the same because of $M$ is of course not the Hamiltonian in the non-relativistic world. It will be important to keep this in mind for the later sections.   

\newpage

\section{Null contractions}
We will now present a different contraction of the Poincare algebra. This is inequivalent to the one described earlier as this will be done in light-cone coordinates which are not related to inertial frames by Lorentz transformations. We will look at two cases, one where we focus on one of the null directions and null gradients become dominant and one where we focus on the transverse directions, thus making spatial gradients important.  

\subsection{Focussing on the null direction}\label{car-con}
We will for the moment work in four spacetime dimensions labelled by $(t, x,y,z)$ and define the lightcone coordinates as
\be{}
x^\pm= \frac{1}{\sqrt{2}}(t\pm z). 
\ee
Hence the derivatives are given by $\d_\pm= \frac{1}{\sqrt{2}} (\d_t\pm\d_z)$. We redefine the Poincare generators as follows:
\begin{subequations}\label{null-poi}
\bea{}
&& P_+ =\frac{1}{\sqrt{2}}( P_t + P_z) = \d_+, \quad P_- = \frac{1}{\sqrt{2}}(P_t - P_z )= \d_-, \\
&& J^{1}_i =\frac{1}{\sqrt{2}}(J_{it} - J_{iz}) =  x^+\d_i + x_i\d_-, \quad J^{2}_i = \frac{1}{\sqrt{2}}(J_{it} + J_{iz})  =  x^-\d_i + x_i\d_+, \\
&& P_i = \d_i, \quad J_{ij} = x_i\d_j - x_j\d_i, \quad B= J_{zt} = x^+\d_+ - x^-\d_-.
\eea
\end{subequations}
\begin{center}
\begin{figure}[t]
\includegraphics[scale=0.5]{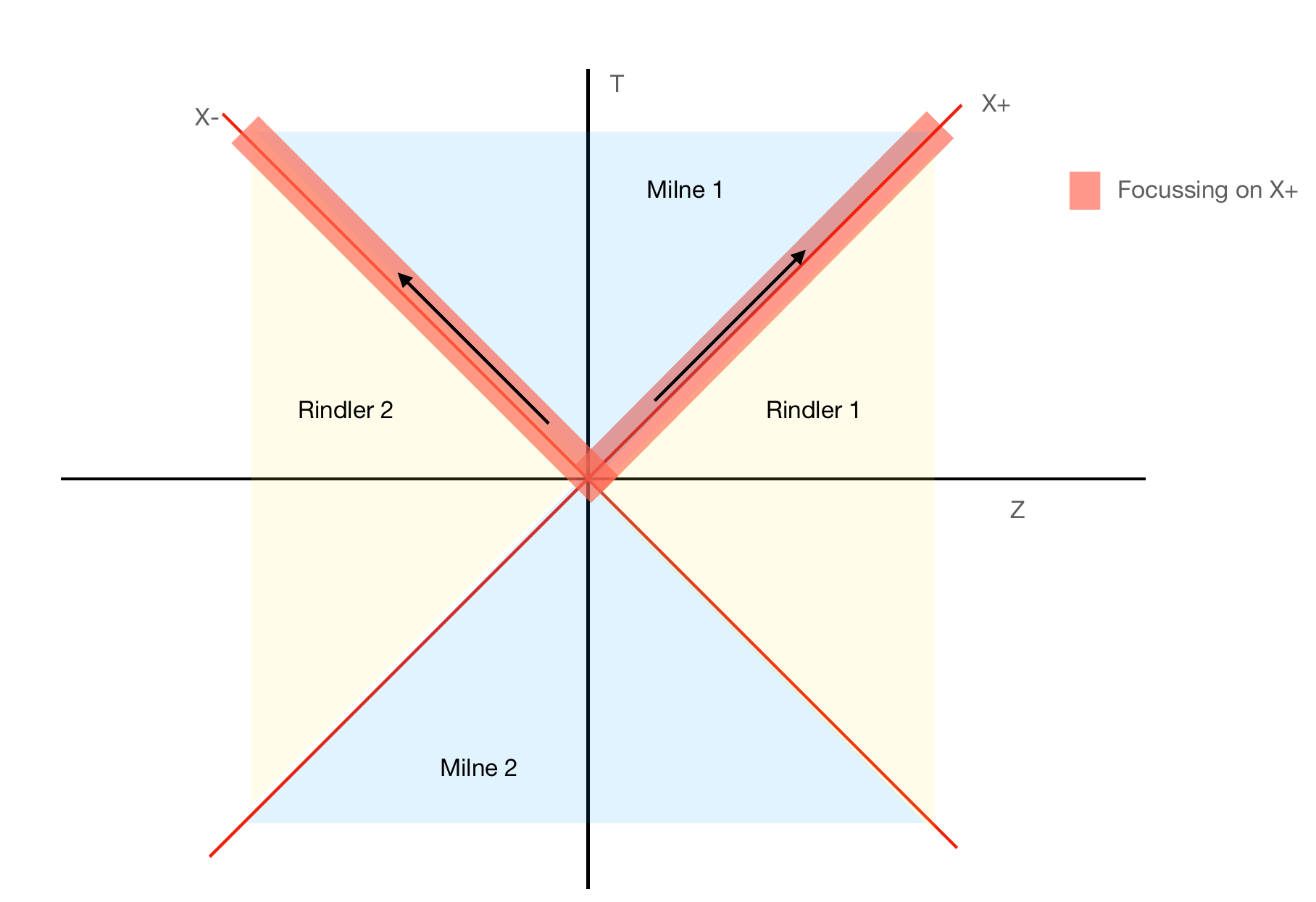}
\caption{Contraction of null direction: The symmetric version}
\label{fig1}
\end{figure}
\end{center}
We now are interested in focussing on focussing on the null direction $x^-$. So the contraction we will do is
\be{cont-}
x^+ \to x^+, \, x^- \to \e x^-, \, x^i \to x^i, \quad \e \to 0.
\ee 
We could have of course focussed on $x^+$ instead. The results would be unchanged. We will have more to say about this below. Now with this contraction \refb{cont-}, the generators become
\begin{subequations}\label{2carr}
\bea{}
&& H_+=\lim_{\e\to 0} P_+= \d_+, \quad C_{+i} = \lim_{\e\to 0} J^{2}_i = x_i\d_+, \\
&& H_-= \lim_{\e\to 0} \e P_-=\d_-, \quad C_{-i} = \lim_{\e\to 0} \e J^{1}_i = x_i\d_-,  \\
&& P_i = \d_i, \quad J_{ij} = x_i\d_j - x_j\d_i, \quad B = x^+\d_+ - x^-\d_-.
\eea
\end{subequations}

It is very interesting to note that now we get {\em two co-dimension one overlapping Carrollian algebras} in the limit or two 3d Carrollian algebras. The subalgebras are 
\begin{subequations}
\begin{align}
\text{Carroll(+):} & \quad \{H_+, C_{+i}, P_i, J_{ij} \}, \\
\text{Carroll(--):} & \quad \{H_-, C_{-i}, P_i, J_{ij}\}. 
\end{align}
\end{subequations}
The generator $B$, which is the original $(tz)$ boost operator, is also a very interesting one. It has the property
\be{lc-boost}
[\O_\pm, B] = \pm \O_\pm, \quad [\O_T, B] = 0, 
\ee
where $\O_\pm = (H_\pm, C_{i\pm})$ are the operators which are on the lightcone and $\O_T= (P_i, J_{ij})$ are the transverse operators. It is important to note that the $\O_+$ and $\O_-$ operators have zero cross commutation relations and hence do not talk to each other at all in the algebra. These are thus disjoint sectors in the algebra and hence of any system that has this algebra as a symmetry. The sectors corresponding to motion along $x^+$ or $x^-$. $x^\pm$ become the time directions of the Carrollian observer on each individual null boundary. These sectors correspond to the observers for whom there is no access to the negative time direction. This would e.g. be true for the observer living in the top Milne patch, as depicted in Fig.\ref{fig1} above. We should emphasise that the other contraction $x^+\to \e x^+$ would have lead to an identical contraction. 
\begin{center}
\begin{figure}[h]
\includegraphics[scale=0.5]{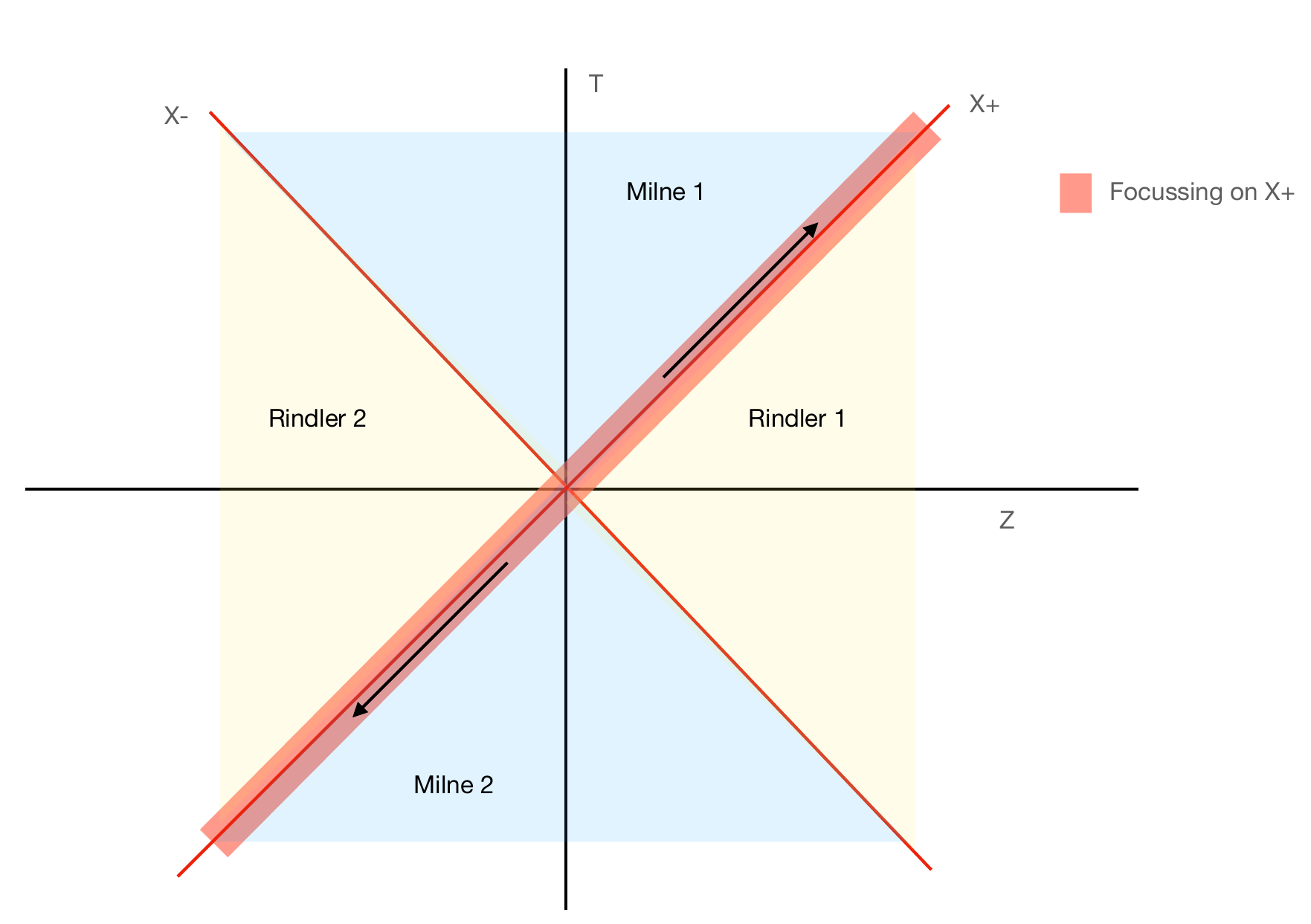}
\caption{Contraction of the null direction: the chiral/anti-chiral picture}
\label{fig2}
\end{figure}
\end{center}
\subsection*{A different basis: the chiral-antichiral picture}
Let us make some other comments here. If we did not recombine the original Poincare generators as \refb{null-poi}, but just chose to work in the original form of the generators \refb{poincare} just rewritten in the lightcone variables, we would again recover the two overlapping lower dimensional Carrollian algebras, but now instead of having a $(+)$ and a $(-)$ algebras, confined to the $x^+$ and $x^-$ axes, we would have only modes moving along one of the null axes, but in either direction. This is straightforward to see, but let us elaborate.
\begin{subequations}
\bea{}
&&P_0 = \d_t = \frac{1}{\sqrt{2}}(\d_+ + \d_-) \to H_1= \d_ - , \, P_z = \d_z = \frac{1}{\sqrt{2}}(\d_+ - \d_-) \to H_2 = - \d_ -, \\
&&J_{0i} = x_i\d_t + t\d_i = \frac{1}{\sqrt{2}} \left(x_i( \d_+ + \d_-) + (x_+ + x_-)\d_i\right) \to C^1_i= x_i \d_-, \\
&& J_{iz} = x_i\d_z - z\d_i =  \frac{1}{\sqrt{2}} \left( x_i( \d_+ - \d_-) + (x_+ - x_-)\d_i \right)\to C^2_i= -x_i \d_-
\eea
\end{subequations}
In the above, the rightarrow ``$\to$'' indicates the place where the contraction has been made. 
We again see that $\{H_1, C^1_{i}, P_i, J_{ij} \}$ and $\{H_2, C^2_{i}, P_i, J_{ij}\}$ form two overlapping 3d Carroll algebras. $H_1$ corresponds to motion forward in the null time direction, while $H_2$ corresponds to backward motion in the time direction. We depict this in Fig.\ref{fig2} above. 

\subsection*{The main message}
By choosing coordinates appropriately, we can focus on observers in Rindler and Milne patches of the Minkowski spacetime. But the overall message stays the same: 

\smallskip

Contraction of the Poincare algebra {\em in lightcone coordinates} along one of the {\em lightlike directions} gives rise to {\em two copies} of {\em co-dimension one Carrollian algebras} each corresponding to motion along one of the lightlike directions.

\smallskip

\subsection{Focussing on the transverse direction}\label{gal-con}
We now focus on the spatial or non-null directions in the theory. This means we will consider a case where the spatial derivatives are large compared to the null ones. This will be achieved by the ``transverse'' contraction:
\be{}
x_i \to \e x_i, \quad x^\pm \to x^\pm, \quad \e\to0. 
\ee
We will now consider the effect of this contraction on the generators \refb{null-poi}: 
\bea{2gal}
&& H_+= \d_+, \quad G^-_{i} = \lim_{\e\to 0} \e J^{2}_i = x^-\d_i, \quad  H_-= \d_-, \quad G^+_{i} = \lim_{\e\to 0} \e J^{1}_i = x^+\d_i,  \\
&& P_i = \d_i, \quad J_{ij} = x_i\d_j - x_j\d_i, \quad B = x^+\d_+ - x^-\d_-.
\eea

The contraction in this case gives us two overlapping co-dimension one {\em Galilean} algebras. The generators forming the subalgebras are 
\begin{subequations}\label{g+-}
\begin{align}
\text{Galilei(+):} & \quad \{H_+, G^+_{i}, P_i, J_{ij} \}, \\
\text{Galilei(--):} & \quad \{H_-, G^-_{i}, P_i, J_{ij}\}. 
\end{align}
\end{subequations}
The lightcone boost $B$ again acts in an identical manner to \refb{lc-boost}. It is interesting to note that in this transverse contraction, one can further extend the Galilei algebras to their Bargmann versions by including the other Hamiltonian as the mass, i.e. 
\begin{subequations}\label{b+-}
\begin{align}
\text{Bargmann(+):} & \quad \{H_+, G^+_{i}, P_i, J_{ij}, M_+=H_- \}, \\
\text{Bargmann(--):} & \quad \{H_-, G^-_{i}, P_i, J_{ij}, M_-=H_+\}. 
\end{align}
\end{subequations}
This of course does not follow from the contraction of the Poincare algebra, so does not hold at the level of the vector fields we have introduced here, but can be added as an extension of the algebra. For the Bargmann $(\pm)$ algebras, the Hamiltonians $H_\mp$ act as central elements, commuting with all generators.

\subsection{Lightcone Poincare and sub-algebras}
We had written down the Poincare generators in the lightcone coordinates in \refb{null-poi}, but not the algebra. It is very instructive to write down these commutators explicitly. 
\begin{subequations}\label{poi+-}
\bea{}
&& \hspace{-0.5cm} [J^{1}_i, J^{2}_j]= B \delta_{ij} + J_{ij}, \, [J^{1, 2}_i, J^{1, 2}_j]= 0, \, [J^{1,2}_i, J_{jk}] = \delta_{i[j} J^{1,2}_{k]}, \, [J_i^{1,2}, P_j] = \delta_{ij} P_\mp, \\
&& [P_+, J^{1}_i] = [P_-, J^{2}_i]= P_i, \, [J^{1}_i, P_-] = [J^{2}_i, P_+]= 0, \,  [J_{ij}, P_{k}] = \delta_{k[j} P_{i]}, \\
&& [J_{ij}, P_{\pm}]=[B, P_i]=[B, J_{ij}]= 0, \, [P_\pm, B]=\pm P_\pm, \, [B, J^{1,2}] = \pm J^{1,2}. 
\eea
\end{subequations}

\paragraph{Carroll sub-algebras:} Now notice the following subalgebra: 
\begin{subequations}
\bea{}
&& [J^{1}_i, J^{1}_j]= 0, \quad  [J_i^1, P_j] = - \delta_{ij} P_- , \quad  [J^{1}_i, P_-] = 0, \quad [J_{ij}, P_{-}] = 0 \\
&& [J^{1}_i, J_{jk}] = \delta_{ij} J^{1}_k - \delta_{ik} J^{1}_j, \quad  [P_{k}, J_{ij}] = \delta_{ik} P_j  - \delta_{jk} P_{i}.
\eea
\end{subequations}
This is isomorphic to \refb{carr} and hence is a 3d Carroll sub-algebra in the 4d Poincare algebra, with $P_-$ playing the role of the Hamiltonian and $J^1_i$ identified as the commuting Carroll boosts. The identifying feature of the Carroll algebra is \refb{car-def}, i.e. the commutator of boost and momenta close to form the Hamiltonian. We see here that
\be{}
[J_i^1, P_j] = - \delta_{ij} P_-. 
\ee
Equivalently, we can also focus on another subalgebra of \refb{null-poi}:
\begin{subequations}
\bea{}
&& [J^{2}_i, J^{2}_j]= 0, \quad  [J_i^2, P_j] = -\delta_{ij} P_+, \quad  [J^{2}_i, P_+] = 0, \quad [J_{ij}, P_{+}] = 0\\
&& [J^{2}_i, J_{jk}] = \delta_{ij} J^{2}_k - \delta_{ik} J^{2}_j, \quad  [ P_{k}, J_{ij}] = \delta_{ik} P_j  - \delta_{jk} P_{i}.
\eea
\end{subequations}
Here $P_+, J^{2}_i$ are the Carroll Hamiltonian and the Carroll boosts. 

\medskip

Crucially, in the full algebra, these subalgebras are not disjoint, by which we mean that the commutators between them are non-zero:
\be{}
[J^{1}_i, J^{2}_j]= B \delta_{ij} +J_{ij}, \quad [P_+, J^{1}_i] = [P_-, J^{2}_i,]= P_i. 
\ee
So, the two overlapping 3d Carrollian sub-algebra reside in 4d Poincare and are given by the sets of generators: $\{P_-, J^1_{i}, P_i, J_{ij} \}$ and $\{P_+, J^2_{i}, P_i, J_{ij}\}$. But these have non-zero cross commutators. 

\medskip

In Sec \ref{car-con}, we found disjoint 3d Carrollian algebras in the limit. Hence the effect of the contraction is to get rid of these cross commutators between the generators. It is instructive to show this explicitly: 
\bea{}
&& C_{-i}= \e J^{1}_i, \, H_- = \e P_-,  C_{+i}= J^{2}_i, \, H_+ = P_+, \\
&& \left[J^1_i, J^2_j\right] = \left[\frac{C_{-i}}{\e}, C_{+j}\right]= B \delta_{ij} +J_{ij}\Rightarrow [C_{-i}, C_{+j}] = \e \left(B \delta_{ij} +J_{ij}\right)\to 0. \\
&& \left[P_+, J^1_i \right] = \left[H_+, \frac{C_{-i}}{\e} \right]= P_i \Rightarrow [H_+, C_{-i}] = \e P_i \to 0. 
\eea

The algebras on the two null surfaces become independent. This is what is achieved by zooming into one of the null directions. 

\medskip

\paragraph{Galilei sub-algebras:} Now consider a different subalgebra: 
\begin{subequations}
\bea{}
&& [J^{1}_i, J^{1}_j]= 0, \quad  [P_+, J^{1}_i] = P_i,  \quad [J_i^1, P_j] = -\delta_{ij} P_-, \quad [J_{ij}, P_{+}] = 0, \\
&& [J^{1}_i, J_{jk}] = \delta_{ij} J^{1}_k - \delta_{ik} J^{1}_j, \quad  [P_{k}, J_{ij}] = \delta_{ik} P_j  - \delta_{jk} P_{i}, \quad [P_-, \text{anything}] = 0.
\eea
\end{subequations}
This set of generators $\{P_+, J^1_i, P_i, J_{ij}, P_-\}$ is a 3d Bargmann sub-algebra of the 4d Poincare algebra, with $P_+$ identified with the Hamiltonian, $P_-$ with the mass (or number) operator and $J^1_i$ the Galilean boosts. Here the defining relation is \refb{gal-def}, i.e. the commutator of Hamiltonian and boosts close to the momenta, which is the case here. 
\be{}
[P_+, J^{1}_i] = P_i
\ee
Additionally, we also have the equivalent of the Bargmann extension \refb{barg}
\be{}
[J_i^1, P_j] = -\delta_{ij} P_-
\ee

Like in the Carroll case, we have another equivalent subalgebra of \refb{null-poi} which again closes to form the Bargmann algebra:
\begin{subequations}
\bea{}
&& [J^{2}_i, J^{2}_j]= 0, \quad  [P_-, J^{2}_i] = P_i, \quad [J_i^2, P_j] = - P_+ \delta_{ij}, \quad [J_{ij}, P_{-}] = 0, \\
&& [J^{2}_i, J_{jk}] = \delta_{ij} J^{2}_k - \delta_{ik} J^{2}_j, \quad  [P_{k}, J_{ij}] = \delta_{ik} P_j  - \delta_{jk} P_{i} \quad [P_+, \text{anything}] = 0
\eea
\end{subequations}
Again, in the full algebra, these two Bargmann subalgebras are not disjoint, by which we mean that the commutators between them are non-zero:
\be{add-g}
[J^{1}_i, J^{2}_j]= B \delta_{ij} +J_{ij}, \quad \left[J^{1,2}_i, P_j \right] = - \delta_{ij} P_\mp
\ee

Like in the Carroll case, in Sec \ref{gal-con}, we found two disjoint Galilean sub-algebras in the limit. A procedure very similar to the one in the Carroll case gets rid of the cross commutation. Let us see how.  We first remind the reader of how the generators scale: 
\be{sc-g}
H_+= P_+, G_{+i} = \e J^{1}_i, H_-= P_-, G_{-i} = \e J^{2}_i,  P_i \to \e P_i, J_{ij} \to J_{ij}, B\to B
\ee
Now the algebra contracts as follows: 
\begin{subequations}
\bea{}
&& \left[J^1_i, J^2_j\right] = \left[\frac{G_{+i}}{\e}, \frac{G_{-j}}{\e}\right]= B \delta_{ij} +J_{ij} \Rightarrow [G_{+i}, G_{-j}] = \e^2 \left(B \delta_{ij} +J_{ij}\right)\to 0. \\
&& \left[J^{1,2}_i, P_j \right] = \left[\frac{G_{\pm i}}{\e}, \frac{P_j}{\e} \right]= - \delta_{ij} P_\mp \Rightarrow [G_{\pm i}, P_j] = - \e^2  \delta_{ij} P_\mp \to  0.
\eea
\end{subequations}
So we see, on top of decoupling the (+) and (-) algebras, the transverse contraction also gets rid of the mass term, as is expected from a non-relativistic contraction of the Poincare algebra. If we impose an ad hoc scaling 
\be{}
M_\pm= \e^2 P_\mp
\ee
to the algebra, not motivated by the spacetime realisation, then we can keep the Bargmann extension in the contraction as well. 

\subsection{Discussions}
The existence of a lower dimensional Galilean sub-algebra with a mass extension to a Bargmann algebra was noticed long ago by Susskind in \cite{Susskind:1967rg}. This formed the basis of what has come to be known as the infinite momentum frame (earlier introduced by Weinberg in \cite{Weinberg:1966jm}) and also is central to the ideas of Discrete Light Cone Quantization (DLCQ), especially in the context of Matrix theory \cite{Taylor:2001vb} (the BFSS matrix model \cite{Banks:1996vh}) and related to more recent ideas of null reductions.

\medskip

We have shown that there is an analogous construction for lower dimensional Carrollian subalgebras inside the Poincare algebra written in lightcone coordinates. We have also emphasised that there are two Galilean/Carrollian algebras corresponding to the propagations along $x^\pm$ respectively. 

\medskip

In the initial works, it was not very clear (at least to us) as to how one would focus on the Galilean subalgebra of the Poincare algebra. With our limiting procedure, we have made this precise. We have shown that the {\em transverse} limit, which makes the spatial derivatives large (as is expected in the infinite momentum frame) effectively disentangles the two Galilean algebras and depending on what one chooses as the time axis, one gets Galilean physics. Similarly, the {\em longitudinal} contraction, which makes the null derivatives important, separates the two Carroll sub-algebras and Carrollian physics on the co-dimension one null surface emerges naturally.

\section{Conformal null contractions}
We will now extend our analysis to the conformal algebras and examine results for the longitudinal and transverse contractions. The expectation is that we would obtain two co-dimension one conformal Carrollian/Galilean algebras from the relativistic conformal algebra. While the Carroll case agrees with expectations, the Galilean case is somewhat counter-intuitive, as we see below. 

\subsection{Usual contractions of the conformal algebra}
In this section, we will revisit the ultra relativistic/ non-relativistic limit of the finite-dimensional conformal group. Here, we have  extra generators of dilatations ($D$) and special conformal transformations ($K_\mu$) on top of the usual Poincare generators \refb{poincare}:
\bea{}
D= x^\mu\d_\mu,\quad K_\mu=2 x_\mu x^\nu\d_\nu-x^2\d_\mu. \label{conformal}
\eea
These generators together with the Poincare generators close to form the relativistic conformal algebra. Along with the Poincare subalgebra \refb{}, the additional commutators that define the algebra are given by: 
\begin{subequations}
\bea{}
&& [D, P_\mu]= - P_\mu,\quad  [D,K_\mu] = K_\mu, \quad [D,J_{\mu\nu}]=0, \\
&& [K_\mu,P_\nu]=-2(J_{\mu\nu}+\eta_{\mu\nu}D) , \quad [K_\lambda,J_{\mu\nu}]=-\eta_{\nu\lambda}K_\mu+\eta_{\mu\lambda} K_\nu, 
\eea
\end{subequations}
We will now look at different contractions of the relativistic conformal algebra.

\paragraph{Carrollian Conformal Algebra:}
We begin by imposing the Carroll limit \refb{carr-limit} on \refb{conformal}. This yields the following set of additional generators along with the usual Carrollian generators \refb{carr-gen}: 
\bea{}
D=t\d_t+x^i\d_i, \quad K= -x^ix_i \d_t, \quad K_i=  2x_ix^\nu\d_\nu -x^kx_k\d_i
\eea
The additional commutation relations are given by
\begin{subequations}\label{ccarr-t}
\bea{}
&& [D,H]=-H,\quad [D,P_i]=-P_i,\quad [D,K_i]=K_i, \quad [D,K]=K \\
&& [H,K_i]=2C_i,\quad [K_i,P_j]=-2(\delta_{ij}D+J_{ij}),\\
&&[K,P_i]=2 C_i, \quad [C_i,K_j]=-\delta_{ij} K,\quad [J_{ij},K_k]=\delta_{jk}K_i-\delta_{ik} K_j,
\eea
\end{subequations}
The above algebra, along with its Carroll sub-algebra \refb{carr}, is called the Carrollian Conformal Algebra. 

\medskip

This is actually the $z=1$ Carrollian Conformal Algebra. The generic class of $z$-Carrollian conformal algebras arise out of the conformal isometries of flat Carrollian manifolds. Carrollian manifolds \cite{Duval:2014uoa, Duval:2014uva} are defined by a pair $(\tau^\mu,  h_{\mu\nu})$, where $\tau^\mu$ is a nowhere vanishing vector field and $h_{\mu\nu}$ is a degenerate covariant metric with $\tau^\mu  h_{\mu\nu}=0$. The isometry algebra of the flat Carroll manifolds 
\be{}
\mathcal{L}_\xi \tau^\mu = 0, \quad \mathcal{L}_\xi h_{\mu\nu} = 0 
\ee
leads to the Carroll algebra. The conformal isometries of the flat Carroll structure: 
\be{}
\mathcal{L}_\xi \tau^\mu = \lambda_1 \tau^\mu, \quad \mathcal{L}_\xi h_{\mu\nu} = -\lambda_2 h_{\mu\nu}, \quad \frac{\lambda_1}{\lambda_2} = \frac{z}{2}. 
\ee
leads to the above mentioned generic $z$-conformal Carroll algebra. The non-Lorentzian nature of the manifold enable us to scale space and time differently leading to a wider range of $z$ other than $z=1$. {\footnote {Also, the isometries as well as the conformal isometries turn out to be infinite dimension for all dimensions and the conformal isometries are also isomorphic to BMS algebras in one higher dimension when $z=1$. In order to restrict to the finite algebra we are discussing here, there needs to be further constraints, specifically setting the arbitrary functions in the solutions of the vector field $\xi$ to be linear.}

\paragraph{Galilean Conformal Algebra:}
The Galilean limit of the relativistic conformal algebra leads to the following additional Galilean conformal generators: 
\begin{subequations}
\bea{}
D=t\d_t+x^i\d_i, \quad K= -t^2\d_t-2 tx^i\d_i, \quad K_i= t^2 \d_i
\eea
\end{subequations}
The commutation relations of these generators are
\begin{subequations}\label{cgal-re}
\bea{}
&& \hspace{-0.4cm}[K_k,J_{ij}] = \delta_{ik} K_j  - \delta_{jk} K_{i}, \quad [G_i,K_j]=0, \quad[K_i,P_j]=0,\\
&& \hspace{-0.4cm}[H,D]=H,\quad [P_i,D]= P_i,\quad [D,K]=K,\quad [D,K_i]=K_i,\\
&&\hspace{-0.4cm}[K,H]=2D,\quad [K,P_i]= 2 G_i,\quad [G_i,K]=-K_i,\quad [H,K_i]=2G_i
\eea
\end{subequations}
This is along with the Galilean algebra (without the additional Bargmann mass extension) now forms the Galilean Conformal Algebra (GCA) \cite{Bagchi:2009my}. 

\medskip

As before with the Carrollian version, we can describe the algebra in a more geometric sense. The above algebra is the $z=1$ version of the $z$-Galilean conformal algebras, which are conformal isometries defined on flat Newton Cartan manifolds. Newton-Cartan manifolds, in close correlation with Carrollian manifolds, are defined by $(\theta_\mu,  g^{\mu\nu})$, where $\theta_\mu$ is a nowhere vanishing one form (sometimes called the clock form) and $g^{\mu\nu}$ is a degenerate contravariant metric with $g^{\mu\nu}\theta_\mu=0$. The Newton-Cartan and Carrollian manifolds are both fibre bundles and are related to each other by the exchange of base and the fibre.  The Galilean algebra is the isometry of the flat Newton-Cartan manifold: 
\be{}
\mathcal{L}_\zeta \theta_\mu = 0, \quad \mathcal{L}_\zeta g^{\mu\nu} = 0. 
\ee
while $z$-Galilean conformal algebras are the conformal isometries of this structure:
\be{}
\mathcal{L}_\zeta \theta_\mu = \lambda_1 \theta_\mu, \quad \mathcal{L}_\zeta g^{\mu\nu} = -\lambda_2 g^{\mu\nu}, \quad \frac{\lambda_1}{\lambda_2} = \frac{z}{2}. 
\ee
The $z=1$ version is the one that is obtained from a Galilean limit of the relativistic conformal algebra as we have stressed. {\footnote{Again, the degenerate structure in the non-relativistic case, as in the Carrollian case, naturally leads to infinite dimensional isometry and conformal isometry. We again need to restrict to linear functions to get back to the finite algebras described above.}} 

\paragraph{Schr{\"o}dinger Algebra:}
Schr{\"o}dinger algebra \cite{Hagen:1972pd, Niederer:1972zz} is the $z=2$ version of the conformal Galilean algebra, as described above as the conformal isometries of flat Newton-Cartan manifolds. This is the maximal symmetry algebra of the free Schr{\"o}dinger wave operator $S=i\d_t+\frac{1}{2m}\d^2_i$ in d+1 dimensions. The algebra is an extension of the Galilean algebra in \refb{gal}, with two additional generators, a single special conformal transformation $\tilde{K}$ , and a dilatation $\tilde{D}$. The generators of the Schr{\"o}dinger algebra are as follows:
\begin{subequations}\label{sch}
\bea{}
&&H= \d_t, \quad P_i = \d_i, \quad G_i = t \d_i, \quad J_{ij} = x_i\d_j -x_j\d_i,\\
&& \tilde{K}= -tx^i\d_i-t^2 \d_t,\quad \tilde{D}=2 t \d_t+x^i\d_i
\eea
\end{subequations}
The relative factor of 2 in the dilatation operator is due to by the fact that this is the $z=2$ theory and the time-like and space-like coordinates scale differently under dilatations: 
\be{dilatation}
x_i\to\lambda x_i,\quad t\to\lambda^2 t.
\ee
The transformation generated by special conformal generator $\tilde{K}$ (which, unlike the $z=1$ GCA, is just one generator instead of $d$ in $d$-spacetime dimensions)
\be{}
x_i\to \frac{x_i}{(1+\lambda t)},\quad t\to\frac{t}{(1+\lambda t)}.
\ee
Non-zero commutators of the Schr{\"o}dinger algebra, includes Galilean commutation relations \refb{gal}, and the following extra brackets:
\begin{subequations}
    \bea{}
    &&[\tilde{K},P_i]= G_i,\quad [\tilde{K},G_i]=0,\quad [\tilde{D},G_i]= G_i, \quad [\tilde{D},P_i]= -P_i \\
    &&[\tilde{D},\tilde{K}]= 2\tilde{K},\quad [\tilde{K},H]=\tilde{D}, \quad[\tilde{D},H]= -2H.    
    \eea
\end{subequations}

\subsection{Null contractions: Longitudinal}
We now perform null contractions of the relativistic conformal algebra. The first step is to write the conformal generators in the lightcone coordinates like the Poincare case. Here again, we will redefine the conformal generators, in keeping with the Poincare algebra, as follows:
\begin{subequations}\label{null-ccon}
\bea{}
&& D=x^+\partial_+ +x^-\partial_-+ x^i\partial_i, \\
&&K^1= - \frac{1}{\sqrt{2}}(K_t-K_z)=x^kx_k\partial_-+2(x^+)^2\partial_++2x^+x^k\partial_k ,\\
&&K^2=-\frac{1}{\sqrt{2}}(K_t+K_z)=x^kx_k\partial_++2(x^-)^2\partial_-+2x^-x^k\partial_k,\\
&&K_i=2x_i(x^+\partial_++x^-\partial_-+x^i\partial_i)+(2x^+x^--x^jx_j)\partial_i, 
\eea
\end{subequations}
The relativistic conformal algebra written in this basis will contain interesting sub-algebras. But we will discuss this in a later sub-section in order to keep the presentation similar to the Poincare case and so that we can highlight how the two contractions act on the sub-algebras. 

\medskip

We now detail the null contractions and first consider the longitudinal contraction. Let us focus on the null direction $x^-$, by using the contraction \refb{cont-}. The contraction of the other direction $x^+$, like before, yields identical results. The $x^-$ contraction leads to the following set of generators
\bea{ccarr-cont}
&& K_-=\lim_{\e\to 0}\e K^1= x^kx_k\partial_-, \, K_+=\lim_{\e\to 0} K^2= x^kx_k\partial_+,  \, D=x^+\partial_+ +x^-\partial_-+ x^i\partial_i, \nonumber\\
&&\bar{K}_i=\lim_{\e\to 0}K_i =2x_i(x^+\partial_++x^-\partial_-+x^i\partial_i) -x^jx_j\partial_i,
\eea
As we expected we get two co-dimension one overlapping Carrollian Conformal algebras in the limit. The sub algebras are
\begin{subequations}
\begin{align}
\text{Conformal Carroll(+):} & \quad \{H_+, C^+_{i}, P_i, J_{ij},D, K_+,\bar{K}_i \}, \\
\text{Conformal Carroll(--):} & \quad \{H_-, C^-_{i}, P_i, J_{ij},D,K_-,\bar{K}_i\}. 
\end{align}
\end{subequations}
The commutations with the boost $B$ is again along similar lines as \refb{lc-boost}, where we now have $\O_{\pm}=\{H_{\pm},C^i_{\pm}, K_{\pm}\}$ and $\O_T = \{ P_i, J_{ij},D,\bar{K}_i \}$. 
%Now we can try to take the same contraction without recombination as already done in \refb{car-con}. This leads to identical results. 

\begin{comment}
Now we can try to take the same contraction without recombination as already done in \refb{car-con}. We define the following generators,
\begin{subequations}
\bea{}
&&\hspace{-1.2cm}\Tilde{P}_t = \sqrt{2}\d_t = \d_+ + \d_- \to H_1= \d_ - , \quad \Tilde{P}_z = \sqrt{2}\d_z =\d_+ - \d_- \to H_2 = - \d_ - \\
&&\hspace{-1.2cm}\Tilde{J}_{it} =\sqrt{2}( x_i\d_t + t\d_i) = x_i( \d_+ + \d_-) +(x^+ + x^-)\d_i \to C^1_i= x_i \d_-, \\
&& \hspace{-1.2cm}\Tilde{J}_{iz} = \sqrt{2}(x_i\d_z - z\d_i )= x_i( \d_+ - \d_-) -(x^+ - x^-)\d_i \to C^2_i= -x_i \d_-,\\
&&\hspace{-1.2cm}\Tilde{K}_t=-2(x^++x^-)x^i\partial_i-(2(x^+)^2+x^ix_i)\partial_+-(2(x^-)^2+x^ix_i)\partial_- \to K^{1}_-=-x^ix_i\d_- \\
&&\hspace{-1.2cm}\Tilde{K}_z=2(x^+-x^-)x^i\partial_i+(2(x^+)^2-x^ix_i)\partial_+-(2(x^-)^2-x^ix_i)\partial_-\to K^{2}_-=x^ix_i\d_- 
\eea
\end{subequations}
Here, right arrow $``\to"$ indicates that the contraction has been made. It is evident that we again have two overlapping copies of  Carrollian conformal algebras. 
\end{comment}

\subsection{Null contractions: Transverse}\label{gal-ccon}
The conformal null contraction in the longitudinal yielded results that we perhaps were expecting and is a direct generalisation of the Poincare case. We now focus on the transverse contraction and here we will meet a surprise. 

\medskip

We scale the non-null directions. This means we will consider a case where the spatial derivatives are large compared to the null ones. We do this by our usual procedure \refb{}. Consider the effect of this contraction on the generators \refb{null-ccon}: 
\begin{subequations}\label{2gal}
\bea{}
&&D=x^+\partial_+ +x^-\partial_-+ x^i\partial_i, \quad  \Tilde{K}_i=\lim_{\e\to 0}\e K_i= 2x^+x^-\d_i, \\&& K^\pm=\lim_{\e\to 0} K^{1,2}= 2(x^\pm)^2\d_\pm +2x^\pm x^k\d_k. 
\eea
\end{subequations}
The algebra of the contracted generators takes the form below. We again focus just on the conformal generators. The other commutations are as in \refb{}.  
\begin{subequations}\label{calgebra}
\bea{}
&&[H_\pm,D]=H_\pm, \, [P_{i},D]=P_{i},\, [D,G^\pm_i]=0=[D,J_{ij}],\quad [D,K^\pm]=K^\pm,\\
&&[D,B]=0, \quad [D,\Tilde{K}_i]=\Tilde{K}_i,\quad [\Tilde{K}_i,P_j]=0 ,\quad[\Tilde{K}_i,G^{\pm}_{j}]=0,\\
&&[\Tilde{K}_i,J_{jl}]=\delta_{i[j}\Tilde{K}_{l]},\quad [\Tilde{K}_i,K^\pm]=0=[\Tilde{K}_i,\Tilde{K}_j],\\
&&[K^\pm,P_i]=-2 G^\pm_i,\quad [K^+,K^-]=0,\quad [H_\pm,K^\pm]=2(D\pm B),\\
&&[K^\pm,G^\mp_{i}]=-\Tilde{K}_i,\quad [B,K^+]= K^+,\quad [B,K^-]=-K^-,\\
&&[H_\pm,\Tilde{K}_i]=2G^\mp_i,\quad [B,\Tilde{K}_i]=0,\quad [K^\pm,G^\pm_{i}]=0=[K^\pm,J_{ij}].
\eea
\end{subequations}
The algebra above is not a natural analogue of what we found for the transverse null contraction of the Poincare algebra, because there are no lower dimensional GCA subalgebras that emerge here. Notice in particular the commutators:
\be{hk}
[H_+,K^+]=2(D+ B), \quad  [H_-,K^-]=2(D- B)
\ee
These certainly do not obey the commutation required of the GCA, which has
\be{}
[H, K] = 2D.
\ee
However, a hint at the solution already lies in the expression for the RHS of the commutation relations in \refb{hk}. To see this explicitly, let us write this out:
\begin{subequations}
\bea{}
D+B = x^+\partial_+ +x^-\partial_-+ x^i\partial_i + x^+\partial_+ - x^-\partial_- = 2 x^+\partial_+ + x^i\partial_i = \tilde{D}^+ \\
D-B = x^+\partial_+ +x^-\partial_-+ x^i\partial_i - x^+\partial_+ + x^-\partial_- = 2 x^-\partial_- + x^i\partial_i = \tilde{D}^-
\eea
\end{subequations}
So, what we get are two Schr{\"o}dinger dilatation operators with $x^\pm$ acting as the time directions. The analogy is taken to its completion when we identify
\be{}
\Tilde{K}^+=-\frac{1}{2}K^+=-(x^+)^2\d_+-x^+x^k\d_k,\quad \Tilde{K}^-=-\frac{1}{2}K^-=-(x^-)^2\d_--x^-x^k\d_k
\ee
So $\Tilde{K}^\pm$ are the special conformal generators of the two co-dimension one Schr{\"o}dinger algebras which are now given by 
\begin{subequations}
\begin{align}
& \text{Schr{\"o}dinger(+):}  \quad \{H_+, G^+_{i}, P_i, J_{ij}, \tilde{D}^+, \Tilde{K}^+ \}, \\
& \text{Schr{\"o}dinger(--):}  \quad \{H_-, G^-_{i}, P_i, J_{ij}, \tilde{D}^-, \Tilde{K}^- \}. 
\end{align}
\end{subequations}
There is also the Bargmann mass extension which can be added to the above, like the Galilean algebra \refb{b+-}. 

\medskip

However, interestingly, unlike the conformal Carrollian case and the Carroll and Galilei cases before that, these algebras are not disjoint. The special conformal generator from one side has a non-trivial commutator with the boost generator of the other side: 
\be{}
[K^{\pm},G^{\mp}_i]=-\Tilde{K}_i.
\ee
So we see that in the conformal case, the transverse contraction is different from the other cases we have addressed. We do not get two copies of the $z=1$ GCA in one lower dimension as we may have naively expected, but two co-dimension one $z=2$ Schr{\"o}dinger algebras. Also, these Schr{\"o}dinger algebras are not disjoint.

\subsection{Lightcone Conformal and sub-algebras}
Following our procedure in the Poincare case, we now write down the relativistic conformal algebra in lightcone coordinates and identify various interesting subalgebras. We will see that the two conformal Carroll subalgebras we identify are not disjoint in the entire algebra but the contraction makes them disjoint, very like the case of the Poincare algebra and its sub-algebras. We can also identify Schr{\"o}dinger subalgebras, but we show that these cannot be disentangled by the transverse limit. 

\medskip

We begin by explicitly writing down the relativistic conformal algebra in the lightcone variables. The generators are given by \refb{null-poi} and \refb{null-ccon}. Below we show the commutators with the conformal generators. The whole algebra also contains the Poincare sub-algebra \refb{poi+-}, which we omit below.  
\begin{subequations}\label{calgebra}
\bea{}
&&[P_{\pm},D]=P_{\pm}, \quad [P_{i},D]=P_{i}, \quad [D,J^{1,2}_i]=[D,J_{ij}]=[D,B]=0,\, \\
&&[D,K^{1,2}]=K^{1,2}, \quad [D,K_i]=K_i,\quad [K_i,P_j]=0,\quad[K_i,J^{1,2}_{j}]=-\delta_{ij}K^{1,2},\\
&&[K_i,J_{jl}]=\delta_{i[j}K_{l]},\, [K_i,K^{1,2}]=[K_i,K_j]=[K^1,K^2]=0, \, [P_i, K^{1,2}]= 2 J^{1,2}_i\\
&& [P_+,K^1]=2(D+B), \,  [P_-,K^2]=2(D-B) \quad [K^{1,2},J^{2,1}_{i}]=-K_i\\
&& [B,K^1]= K^1,\, [B,K^2]=-K^2, \, [B,K_i]=0,\\ 
&& [P_+,K_i]=2J^2_i,\, [P_-,K_i]=2J^1_i, \quad [K^{1,2},J^{1,2}_{i}]=0=[K^{1,2},J_{ij}].
\eea
\end{subequations}
\paragraph{Carroll conformal sub-algebras:} We can readily identify two co-dimension one Conformal Carroll subalgebras from the above. These are given by: 
\begin{itemize}
\item \textit{Conformal Carroll Copy 1:} Generators $\{ P_-, J^{1}_i, K^{1}, P_i, J_{ij}, D, K_i\}$.  
\item \textit{Conformal Carroll Copy 2:} Generators $\{ P_+, J^{2}_i, K^{2}, P_i, J_{ij}, D, K_i\}$.
\end{itemize}
Similar to the Poincare case, even in the conformal algebra, these subalgebras above have non-zero cross commutators:
\begin{subequations}
\bea{}
&&[J^{1}_i, J^{2}_j]= B \delta_{ij} + J_{ij}, \quad [P_+, J^{1}_i] = [P_-, J^{2}_i,]= P_i, \\
&&[P_+,K^1]=2(D+B), \quad  [P_-,K^2]=2(D-B), \quad[K^{1,2},J^{2,1}_{i}]=-K_i
\eea
\end{subequations}
We now clarify how the effect of the contraction is to get rid of these cross commutators between the generators of the two sub-algebras making them disjoint. The contractions in terms of the generators are:
\be{}
H_+ = P_+, C_{+i}= J^{2}_i, K_+= K^2,  ~~ H_- = \e P_-, C_{-i}= \e J^{1}_i, K_-=\e K^1.
\ee
Additionally, $D, J_{ij}, K_i$ do not scale. It is now straight-forward to see the action on the algebra. We will consider only the part not included in the Carroll subalgebra, i.e. the ones involving the conformal generators: 
\begin{subequations}
\bea{}
&& \left[P_+,K^1\right]=\left[P_+,\frac{K_-}{\e}\right]=2(D+B) \Rightarrow \left[P_+,K_-\right]=2\e(D+B) \to 0, \\
&&\left[P_-,K^2\right]=\left[\frac{P_-}{\e},K_+\right]=2(D-B) \Rightarrow \left[P_-,K_+\right]=2\e(D-B) \to 0, \\
&&\left[K^1,J^2_i\right]=\left[\frac{K_-}{\e},C_{+i}\right]=-K_i \Rightarrow{}\left[K_-,C_{+i}\right]=-\e K_i\to 0,\\
&&\left[K^2,J^1_i\right]=\left[K_+,\frac{C_{-i}}{\e}\right]=-K_i \Rightarrow{}\left[K_+,C_{-i}\right]=-\e K_i\to 0,
\eea
\end{subequations}
We see explicitly that the cross commutators between the two $(\pm)$ algebras vanish, leading to disjoint sub-algebras that don't talk to each other, just like in the Poincare algebra.  

\paragraph{Schr{\"o}dinger sub-algebras:} We also have two codimension one Schr{\"o}dinger algebras in \refb{calgebra}. These are
\begin{itemize}
\item \textit{Schr{\"o}dinger Copy 1:} Generators $\{ P_+, J^{1}_i, P_i, J_{ij}, \tilde{D}^+=D+B, \Tilde{K}^+= \frac{1}{2} K^{1} \}$.  
\item \textit{Schr{\"o}dinger Copy 2:} Generators $\{ P_-, J^{2}_i, P_i, J_{ij}, \tilde{D}^-= D-B, \Tilde{K}^-= \frac{1}{2} K^{2} \}$.
\end{itemize}
Just like the conformal Carroll case, there are non-zero cross commutators. There are the ones involving the Galilean generators given in \refb{add-g}. The one involving the conformal generators is given by: 
%\begin{subequations}
\bea{}
[K^{1,2},J^{2,1}_{i}]=-K_i
\eea
%\end{subequations}
The generators scale according to \refb{sc-g} and additionally:
\be{}
\tilde{D}^\pm = D \pm B, \quad   \Tilde{K}_i=\e K_i, \quad K^\pm= K^{1,2}. 
\ee
Now consider the scaling of the non-zero cross commutator:
\be{}
\left[K^1,J^2_i\right]=\left[{K_+},\frac{G_{-i}}{\e}\right]=-\frac{{\tilde K}_i}{\e} \Rightarrow{}\left[K_+,G_{-i}\right]=- K_i \\
\ee
The other commutator $[K^2,J^1_i]$ follows identically. We see explicitly that the contraction does not help get rid of the cross-commutator and the algebras are not disjoint.

\section{Conclusions and future directions}

\subsection*{Summary}

In this paper, we have considered novel contractions of the Poincare and the relativistic conformal algebras that we call null contractions. These contractions are different from the usual Galilean and Carrollian contractions of the above algebras because of the fact that we have used the lightcone coordinates which are unreachable by a Lorentz transformation from usual inertial coordinates. 

\begin{itemize}

\item {\em Different subalgebras:} Written in lightcone coordinates, the Poincare algebra already contains two Carrollian sub-algebras in one lower dimension as well as two co-dimension one Galilean algebras. The existence of lower dimensional Galilean sub-algebra of the Poincare was famously pointed out by Susskind and as remarked earlier, is central to the discussions of the infinite momentum frame and related physics. One of our new observations is the fact that there are also co-dimension one Carroll subalgebras in the Poincare algebra. This, in hindsight, is not a surprise, given the fact that the $d+1$ dimensional Poincare algebra is isomorphic to the $d$ dimensional conformal Carroll algebra, both being $iso(d,1)$. The Carroll subalgebra is thus obvious. The fact that there exists two such algebras is somewhat curious. It is also curious that the conformal Carroll algebra contains massive Galilean subalgebras in it.

\item {\em Longitudinal null contraction:} The process of contraction makes the picture rather appealing. The longitudinal contraction focusses on the null direction and decouples the two previously interlinked Carroll sub-algebras in the Poincare algebra. This is the limit where the gradients in the null direction become important and we get two Carroll algebras corresponding to the two null directions $x^\pm = \frac{1}{\sqrt{2}}(t\pm z)$. These become the Carrollian time directions for the two null surfaces.

\item {\em Transverse null contraction:} The transverse contraction makes the gradients in the spatial directions large and hence sends the spatial momenta to infinity. We believe this is what makes the infinite momentum frame picture more mathematically precise. The limit also allows us to focus purely on the Galilean algebra without having to neglect some additional non-Galilean commutation relations.

\item {\em Lightlike observers:} As mentioned in the introduction, by looking at these two different null contractions, the longitudinal and the transverse, we can now solve an apparent puzzle. It has been long postulated that observers moving at very high velocities near the speed of light would see the world as non-relativistic or Galilean. This forms a large existing literature on the physics of the infinite momentum frame, which is used in many different aspects of high energy physics, including applications to QCD \cite{Brodsky:1997de} and matrix models in string and M-theory \cite{Taylor:2001vb, Banks:1996vh}. On the other hand, the symmetries associated with null surfaces are Carrollian symmetries and naturally observers which move near lightspeed should see their lightcones close and see Carrollian physics emerging. This apparent contradiction is resolved by understanding that the light-like observer sees both Carrollian and Galilean physics depending on what they focus on. When they are interested in physics of the world they are flying past, they want to focus on the spatial directions and hence spatial gradients become important. This is the physics of the infinite momentum frame which is Galilean. When the lightlike observers wish to focus on what is happening to themselves, they focus on the null direction and hence the gradients along the null directions become important. This leads to the closing up of their lightcones and to Carrollian physics.

\item {\em Conformal considerations:} In the last part of our paper, we focussed on generalising our construction to the relativistic conformal algebra. We found analogous results for the Carroll case, i.e. the relativistic conformal algebra had two lower dimensional conformal Carroll algebras which were entangled to begin with but became disjoint on taking the longitudinal null limit. However, the Galilean case was curiously different. Here there were no $z=1$ Galilean conformal subalgebras. Instead we found two non-disjoint $z=2$ Schr{\"o}dinger algebras which the transverse contraction was not able to separate out. 

\end{itemize}

\subsection*{Future directions}
We hope that this work is the first in a long series of papers that explores null contractions. There are numerous immediate directions of research, some of which are currently under investigation. Below we mention a few of these. 
\begin{itemize}
\item {\em Geometric structure and representation theory:} One of basic things to address is the understanding of this limit in terms of the underlying geometry. This is work in progress. Also, it is important to figure out how the representations of Poincare and relativistic conformal algebras split into the representations of the different underlying sub-algebras and what the null contractions do to the representations. 

\item {\em Generalisation to generic Carroll manifolds:} What we have achieved in this algebraic set up is the projection of bulk symmetries to their null boundaries. This should be doable for generic backgrounds and not just Minkowski. A natural setting in which our analysis can be employed is on the tangent space where similar co-dimension one subalgebras would naturally arise. One can then attempt to ``stitch" together these locally flat patches to construct null surfaces in generically curved backgrounds. 

\item {\em Connections to Penrose limits:} Our algebraic construction is very reminiscent of the Penrose limit of a generic spacetime where one zooms near a null geodesic. It would be good to clarify this point. 

\item {\em Carroll expansions:} One of the important recent tools employed in Carrollian theories is the method of expansion \cite{deBoer:2021jej}. Given a relativistic field theory with a certain action, all fields and parameters are expanded in powers of $c$ and the leading and subleading order terms in the expansion lead to the Electric and Magnetic Carroll theories corresponding to the relativistic theory. In our case, such a procedure should also be straight-forward to implement. One would rewrite the relativistic action in terms of lightcone coordinates and then carry out the expansions. These should lead to two copies of lower dimensional Carroll actions. This is work in progress. 

\item {\em Revisiting the Infinite Momentum Frame:} QFT in the infinite momentum frame \cite{Kogut:1969xa} or equivalently on the lightcone has been used extensively in the context of particle physics and in particular QCD (see e.g. \cite{Brodsky:1997de} for a review). Aspects of quantization on the lightcone treating the null direction as the direction of evolution go back to Dirac's seminal work \cite{Dirac:1949cp}. These discussions were given a more physical footing by Weinberg \cite{Weinberg:1966jm}, who showed that some aspects QFTs become more tractable in the lightcone coordinates. Susskind \cite{Susskind:1967rg} then showed that this simplification was due to the lower dimensional Galilean subalgebra that we have discussed above. These non-relativistic aspects have been explored in the context of particle physics extensively. It would be good to go back to the rather vast literature and see what lessons we can draw about Carrollian physics from here {\footnote{Look at \cite{Barnich:2024aln} for an interesting recent consideration of the Carrollian scalar model and lightcone quantization.}} and in particular what role Carroll has to play in the context of QCD. 

\item {\em Discrete Light Cone Quantization (DLCQ):} DLCQ is the process by which one compactifies the null direction and quantizes the theory in the lightcone. DLCQ is at the heart of the BFSS conjecture \cite{Banks:1996vh} which relates the large $N$ limit of a supersymmetric matrix quantum mechanics to M-theory in lightfront coordinates. In this infinite momentum frame, the dynamics of M-theory become non-relativistic and are effectively captured by a matrix quantum mechanics. Given that now we understand there can be associated Carrollian physics, it would be good to figure out whether there are aspects of M-theory that are governed by Carrollian physics. 

\item{\em Lessons for flat holography?} One of the prominent paths to holography in flat spacetime is the so called Carrollian holography programme which posits that gravitational physics in asymptotically Minkowskian spacetimes is described by a co-dimension one Carroll CFT that lives on the null boundary (see e.g. \cite{Bagchi:2023cen} for a recent discussion). In the first part of our paper, we have seen that when the lightcone is approached, the Poincare algebra splits into two lower dimensional Carroll algebras. Null infinity $\mathscr{I}^\pm$ is a very particular null surface and it is very likely that the emergence of the two 3d Carroll algebras from the 4d Poincare algebra would get enhanced to a whole conformal class of Carroll algebras and hence would lead to two copies of 3d conformal Carroll (or BMS$_4$) algebras arising from the bulk Poincare symmetry. This would be a very natural way to understand the emergence of the two copies of BMS$_4$ arising on the future and past null boundaries $\mathscr{I}^\pm$. 

The point at spatial infinity in our analysis would correspond to the point where $x^+=x^-=0$ and could be approached by a double scaling $x^-\to \e x^-$ and $x^+\to \e x^+$. It is curious to note that the addition scaling $x^+\to \e x^+$ does not change the algebra \refb{2carr}, which were obtained by focussing around $x^-=0$ by $x^-\to \e x^-$. If there is again a conformal enhancement due to the speciality of $i_0$, this would indicate that there is a BMS algebra also arising at spatial infinity, which is in keeping with existing literature. 

A final word about flat holography. The BFSS matrix model was the initial proposal of a hologram of flat space. It seems that there should be something Carrollian about the BFSS matrix theory as well. Given the above connections between Carrollian theories and flat holography, this is certainly a tantalising direction to investigate. 

\end{itemize}

In conclusion, we believe the simple algebraic exercise we have presented in the current paper has far reaching consequences. We hope to report on some of the immediate questions that have arisen from our analysis in the near future. 

\bigskip \bigskip

\subsection*{Acknowledgements}
AB wishes to especially thank Shahin Sheikh-Jabbari, discussions with whom triggered the questions that have led to the project. We are also grateful to Aritra Banerjee, Daniel Grumiller, Sharang Iyer, Kedar Kolekar, Saikat Mondal and Ashish Shukla  and for useful discussions. 

\medskip

AB is partially supported by a Swarnajayanti Fellowship from the Science and Engineering Research Board (SERB) under grant SB/SJF/2019-20/08 and also by SERB grant CRG/2022/006165. AB acknowledges the support received from the Erwin Schr{\"o}dinger Institute, Vienna during the programme ``Carrollian Physics and Holography''  as well as  the warm hospitality of ULB Brussels, TU Wien and University of Vienna during the course of this work. PS is supported by an IIT Kanpur Institute Assistantship.

\bibliographystyle{JHEP}
\bibliography{ref-null}

\providecommand{\href}[2]{#2}\begingroup\raggedright\begin{thebibliography}{10}

\bibitem{Bergshoeff:2022eog}
E.~Bergshoeff, J.~Figueroa-O'Farrill and J.~Gomis, \emph{{A non-lorentzian
  primer}},
  \href{http://dx.doi.org/10.21468/SciPostPhysLectNotes.69}{\emph{SciPost Phys.
  Lect. Notes} {\bfseries 69} (2023) 1},
  [\href{https://arxiv.org/abs/2206.12177}{{\ttfamily 2206.12177}}].

\bibitem{Hartong:2022lsy}
J.~Hartong, N.~A. Obers and G.~Oling, \emph{{Review on Non-Relativistic
  Gravity}}, \href{http://dx.doi.org/10.3389/fphy.2023.1116888}{\emph{Front. in
  Phys.} {\bfseries 11} (2023) 1116888},
  [\href{https://arxiv.org/abs/2212.11309}{{\ttfamily 2212.11309}}].

\bibitem{Oling:2022fft}
G.~Oling and Z.~Yan, \emph{{Aspects of Nonrelativistic Strings}},
  \href{http://dx.doi.org/10.3389/fphy.2022.832271}{\emph{Front. in Phys.}
  {\bfseries 10} (2022) 832271},
  [\href{https://arxiv.org/abs/2202.12698}{{\ttfamily 2202.12698}}].

\bibitem{LBLL}
J.~Levy-Leblond, \emph{{Une nouvelle limite non-relativiste du group de
  Poincare}}, {\emph{Ann.Inst.Henri Poincare} {\bfseries 3} (1965) 1}.

\bibitem{SenGupta:1966qer}
N.~D. Sen~Gupta, \emph{{On an analogue of the Galilei group}},
  \href{http://dx.doi.org/10.1007/BF02740871}{\emph{Nuovo Cim. A} {\bfseries
  44} (1966) 512--517}.

\bibitem{Bidussi:2021nmp}
L.~Bidussi, J.~Hartong, E.~Have, J.~Musaeus and S.~Prohazka, \emph{{Fractons,
  dipole symmetries and curved spacetime}},
  \href{http://dx.doi.org/10.21468/SciPostPhys.12.6.205}{\emph{SciPost Phys.}
  {\bfseries 12} (2022) 205},
  [\href{https://arxiv.org/abs/2111.03668}{{\ttfamily 2111.03668}}].

\bibitem{Bagchi:2022eui}
A.~Bagchi, A.~Banerjee, R.~Basu, M.~Islam and S.~Mondal, \emph{{Magic fermions:
  Carroll and flat bands}},
  \href{http://dx.doi.org/10.1007/JHEP03(2023)227}{\emph{JHEP} {\bfseries 03}
  (2023) 227}, [\href{https://arxiv.org/abs/2211.11640}{{\ttfamily
  2211.11640}}].

\bibitem{Bagchi:2023ysc}
A.~Bagchi, K.~S. Kolekar and A.~Shukla, \emph{{Carrollian Origins of Bjorken
  Flow}}, \href{http://dx.doi.org/10.1103/PhysRevLett.130.241601}{\emph{Phys.
  Rev. Lett.} {\bfseries 130} (2023) 241601},
  [\href{https://arxiv.org/abs/2302.03053}{{\ttfamily 2302.03053}}].

\bibitem{Bagchi:2023rwd}
A.~Bagchi, K.~S. Kolekar, T.~Mandal and A.~Shukla, \emph{{Heavy-ion collisions,
  Gubser flow, and Carroll hydrodynamics}},
  \href{http://dx.doi.org/10.1103/PhysRevD.109.056004}{\emph{Phys. Rev. D}
  {\bfseries 109} (2024) 056004},
  [\href{https://arxiv.org/abs/2310.03167}{{\ttfamily 2310.03167}}].

\bibitem{Donnay:2019jiz}
L.~Donnay and C.~Marteau, \emph{{Carrollian Physics at the Black Hole
  Horizon}}, \href{http://dx.doi.org/10.1088/1361-6382/ab2fd5}{\emph{Class.
  Quant. Grav.} {\bfseries 36} (2019) 165002},
  [\href{https://arxiv.org/abs/1903.09654}{{\ttfamily 1903.09654}}].

\bibitem{deBoer:2021jej}
J.~de~Boer, J.~Hartong, N.~A. Obers, W.~Sybesma and S.~Vandoren, \emph{{Carroll
  Symmetry, Dark Energy and Inflation}},
  \href{http://dx.doi.org/10.3389/fphy.2022.810405}{\emph{Front. in Phys.}
  {\bfseries 10} (2022) 810405},
  [\href{https://arxiv.org/abs/2110.02319}{{\ttfamily 2110.02319}}].

\bibitem{Bagchi:2010zz}
A.~Bagchi, \emph{{Correspondence between Asymptotically Flat Spacetimes and
  Nonrelativistic Conformal Field Theories}},
  \href{http://dx.doi.org/10.1103/PhysRevLett.105.171601}{\emph{Phys. Rev.
  Lett.} {\bfseries 105} (2010) 171601},
  [\href{https://arxiv.org/abs/1006.3354}{{\ttfamily 1006.3354}}].

\bibitem{Bagchi:2012xr}
A.~Bagchi, S.~Detournay, R.~Fareghbal and J.~Sim{\'o}n, \emph{{Holography of 3D
  Flat Cosmological Horizons}},
  \href{http://dx.doi.org/10.1103/PhysRevLett.110.141302}{\emph{Phys. Rev.
  Lett.} {\bfseries 110} (2013) 141302},
  [\href{https://arxiv.org/abs/1208.4372}{{\ttfamily 1208.4372}}].

\bibitem{Bagchi:2016bcd}
A.~Bagchi, R.~Basu, A.~Kakkar and A.~Mehra, \emph{{Flat Holography: Aspects of
  the dual field theory}},
  \href{http://dx.doi.org/10.1007/JHEP12(2016)147}{\emph{JHEP} {\bfseries 12}
  (2016) 147}, [\href{https://arxiv.org/abs/1609.06203}{{\ttfamily
  1609.06203}}].

\bibitem{Bagchi:2022emh}
A.~Bagchi, S.~Banerjee, R.~Basu and S.~Dutta, \emph{{Scattering Amplitudes:
  Celestial and Carrollian}},
  \href{http://dx.doi.org/10.1103/PhysRevLett.128.241601}{\emph{Phys. Rev.
  Lett.} {\bfseries 128} (2022) 241601},
  [\href{https://arxiv.org/abs/2202.08438}{{\ttfamily 2202.08438}}].

\bibitem{Donnay:2022aba}
L.~Donnay, A.~Fiorucci, Y.~Herfray and R.~Ruzziconi, \emph{{Carrollian
  Perspective on Celestial Holography}},
  \href{http://dx.doi.org/10.1103/PhysRevLett.129.071602}{\emph{Phys. Rev.
  Lett.} {\bfseries 129} (2022) 071602},
  [\href{https://arxiv.org/abs/2202.04702}{{\ttfamily 2202.04702}}].

\bibitem{Bagchi:2013bga}
A.~Bagchi, \emph{{Tensionless Strings and Galilean Conformal Algebra}},
  \href{http://dx.doi.org/10.1007/JHEP05(2013)141}{\emph{JHEP} {\bfseries 05}
  (2013) 141}, [\href{https://arxiv.org/abs/1303.0291}{{\ttfamily 1303.0291}}].

\bibitem{Bagchi:2015nca}
A.~Bagchi, S.~Chakrabortty and P.~Parekh, \emph{{Tensionless Strings from
  Worldsheet Symmetries}},
  \href{http://dx.doi.org/10.1007/JHEP01(2016)158}{\emph{JHEP} {\bfseries 01}
  (2016) 158}, [\href{https://arxiv.org/abs/1507.04361}{{\ttfamily
  1507.04361}}].

\bibitem{Bagchi:2020fpr}
A.~Bagchi, A.~Banerjee, S.~Chakrabortty, S.~Dutta and P.~Parekh, \emph{{A tale
  of three \textemdash{} tensionless strings and vacuum structure}},
  \href{http://dx.doi.org/10.1007/JHEP04(2020)061}{\emph{JHEP} {\bfseries 04}
  (2020) 061}, [\href{https://arxiv.org/abs/2001.00354}{{\ttfamily
  2001.00354}}].

\bibitem{Bagchi:2023cfp}
A.~Bagchi, A.~Banerjee, J.~Hartong, E.~Have, K.~S. Kolekar and M.~Mandlik,
  \emph{{Strings near black holes are Carrollian}},
  \href{https://arxiv.org/abs/2312.14240}{{\ttfamily 2312.14240}}.

\bibitem{Bagchi:2022iqb}
A.~Bagchi, D.~Grumiller and M.~M. Sheikh-Jabbari, \emph{{Horizon strings as 3D
  black hole microstates}},
  \href{http://dx.doi.org/10.21468/SciPostPhys.15.5.210}{\emph{SciPost Phys.}
  {\bfseries 15} (2023) 210},
  [\href{https://arxiv.org/abs/2210.10794}{{\ttfamily 2210.10794}}].

\bibitem{Susskind:1967rg}
L.~Susskind, \emph{{Model of selfinduced strong interactions}},
  \href{http://dx.doi.org/10.1103/PhysRev.165.1535}{\emph{Phys. Rev.}
  {\bfseries 165} (1968) 1535--1546}.

\bibitem{Kogut:1971zd}
J.~B. Kogut, \emph{{QUANTUM ELECTRODYNAMICS AT INFINITE MOMENTUM: APPLICATIONS
  TO HIGH-ENERGY SCATTERING}},  other thesis, 9, 1971.

\bibitem{Majumdar:2024rxg}
S.~Majumdar, \emph{{On the Carrollian Nature of the Light Front}},
  \href{https://arxiv.org/abs/2406.10353}{{\ttfamily 2406.10353}}.

\bibitem{Weinberg:1966jm}
S.~Weinberg, \emph{{Dynamics at infinite momentum}},
  \href{http://dx.doi.org/10.1103/PhysRev.150.1313}{\emph{Phys. Rev.}
  {\bfseries 150} (1966) 1313--1318}.

\bibitem{Taylor:2001vb}
W.~Taylor, \emph{{M(atrix) Theory: Matrix Quantum Mechanics as a Fundamental
  Theory}}, \href{http://dx.doi.org/10.1103/RevModPhys.73.419}{\emph{Rev. Mod.
  Phys.} {\bfseries 73} (2001) 419--462},
  [\href{https://arxiv.org/abs/hep-th/0101126}{{\ttfamily hep-th/0101126}}].

\bibitem{Banks:1996vh}
T.~Banks, W.~Fischler, S.~H. Shenker and L.~Susskind, \emph{{M theory as a
  matrix model: A conjecture}},
  \href{http://dx.doi.org/10.1201/9781482268737-37}{\emph{Phys. Rev. D}
  {\bfseries 55} (1997) 5112--5128},
  [\href{https://arxiv.org/abs/hep-th/9610043}{{\ttfamily hep-th/9610043}}].

\bibitem{Duval:2014uoa}
C.~Duval, G.~W. Gibbons, P.~A. Horvathy and P.~M. Zhang, \emph{{Carroll versus
  Newton and Galilei: two dual non-Einsteinian concepts of time}},
  \href{http://dx.doi.org/10.1088/0264-9381/31/8/085016}{\emph{Class. Quant.
  Grav.} {\bfseries 31} (2014) 085016},
  [\href{https://arxiv.org/abs/1402.0657}{{\ttfamily 1402.0657}}].

\bibitem{Duval:2014uva}
C.~Duval, G.~W. Gibbons and P.~A. Horvathy, \emph{{Conformal Carroll groups and
  BMS symmetry}},
  \href{http://dx.doi.org/10.1088/0264-9381/31/9/092001}{\emph{Class. Quant.
  Grav.} {\bfseries 31} (2014) 092001},
  [\href{https://arxiv.org/abs/1402.5894}{{\ttfamily 1402.5894}}].

\bibitem{Bagchi:2009my}
A.~Bagchi and R.~Gopakumar, \emph{{Galilean Conformal Algebras and AdS/CFT}},
  \href{http://dx.doi.org/10.1088/1126-6708/2009/07/037}{\emph{JHEP} {\bfseries
  07} (2009) 037}, [\href{https://arxiv.org/abs/0902.1385}{{\ttfamily
  0902.1385}}].

\bibitem{Hagen:1972pd}
C.~R. Hagen, \emph{{Scale and conformal transformations in galilean-covariant
  field theory}}, \href{http://dx.doi.org/10.1103/PhysRevD.5.377}{\emph{Phys.
  Rev. D} {\bfseries 5} (1972) 377--388}.

\bibitem{Niederer:1972zz}
U.~Niederer, \emph{{The maximal kinematical invariance group of the free
  Schrodinger equation.}},
  \href{http://dx.doi.org/10.5169/seals-114417}{\emph{Helv. Phys. Acta}
  {\bfseries 45} (1972) 802--810}.

\bibitem{Brodsky:1997de}
S.~J. Brodsky, H.-C. Pauli and S.~S. Pinsky, \emph{{Quantum chromodynamics and
  other field theories on the light cone}},
  \href{http://dx.doi.org/10.1016/S0370-1573(97)00089-6}{\emph{Phys. Rept.}
  {\bfseries 301} (1998) 299--486},
  [\href{https://arxiv.org/abs/hep-ph/9705477}{{\ttfamily hep-ph/9705477}}].

\bibitem{Kogut:1969xa}
J.~B. Kogut and D.~E. Soper, \emph{{Quantum Electrodynamics in the Infinite
  Momentum Frame}},
  \href{http://dx.doi.org/10.1103/PhysRevD.1.2901}{\emph{Phys. Rev. D}
  {\bfseries 1} (1970) 2901--2913}.

\bibitem{Dirac:1949cp}
P.~A.~M. Dirac, \emph{{Forms of Relativistic Dynamics}},
  \href{http://dx.doi.org/10.1103/RevModPhys.21.392}{\emph{Rev. Mod. Phys.}
  {\bfseries 21} (1949) 392--399}.

\bibitem{Barnich:2024aln}
G.~Barnich, S.~Majumdar, S.~Speziale and W.-D. Tan, \emph{{Lessons from
  discrete light-cone quantization for physics at null infinity: bosons in two
  dimensions}}, \href{http://dx.doi.org/10.1007/JHEP05(2024)326}{\emph{JHEP}
  {\bfseries 05} (2024) 326},
  [\href{https://arxiv.org/abs/2401.14873}{{\ttfamily 2401.14873}}].

\bibitem{Bagchi:2023cen}
A.~Bagchi, P.~Dhivakar and S.~Dutta, \emph{{Holography in Flat Spacetimes: the
  case for Carroll}},  \href{https://arxiv.org/abs/2311.11246}{{\ttfamily
  2311.11246}}.

\end{thebibliography}\endgroup

\end{document}